# THE ELECTRON PARAMAGNETIC RESONANCE IN THE STUDY OF TISSUES SPECIMENS

IRENEUSZ STEFANIUK[1]*, DAGMARA WRÓBEL[2], ANDRZEJ SKRĘT[3],
JOANNA SKRĘT – MAGIERŁO[3], TOMASZ GÓRA[4], PIOTR SZCZERBA[4]

[1] Centre for Microelectronics and Nanotechnology, University of Rzeszów, Pigonia 1, 35-959 Rzeszów, Poland
[2] Institute of Nuclear Physics, Polish Academy of Science, Krakow, Poland
[3] Institute of Obstetrics and Medical Lifesaving, ul. Pigonia 6, 35-310 Rzeszów, Poland
[4] Clinical Department of Gynaecology and Obstetrics, County Hospital no.1, FryderykaSzopena2 Rzeszów, Poland

**The Electron Paramagnetic Spectroscopy (EPR) is the most direct and powerful method for the detection and identification of free radicals and other species with unpaired electrons. Statistics disorders are a common gynaecological disorder occurring in women. The condition afflicts around 15% of women to the extent of impairing the quality of living. According to scientific reports as many as 50% of women experiencing problems related to genital statistics disorders. The aim of this work was to investigate tissue taken from women with genital statistics disorders using the Electron Paramagnetic Resonance method. The studies on the tissue of women is one of the first studies in this area. In this work we observed a close relationship between the observed EPR signal and the consumption of omega 3 acids.**

## INTRODUCTION

Electron Paramagnetic Spectroscopy (EPR) is the most direct and powerful method for the detection and identification of free radicals and other species with unpaired electrons (Jackson *et al.*, 2001).This is a method that has been explored previously in considerable detail, particularly in the 1970 and 1980's and there is little evidence in the reference list that the authors have consulted this earlier work, as little of it is referenced (Swartz *et al.*, 1972). Some acknowledgement of these prior studies should be included. More specifically, a number of studies have been carried out on cervical cancer samples using EPR that are of direct relevance to the current study (Wickens *et al.*, 1990). The methods used to process the samples are rather vague and these need to be clarified as a number of studies have provided evidence that some methods of processing human (and animal tissues) for EPR introduce artefacts (Shuter *et al.*, 1990). The EPR investigations have on the aim diagnosis of the paramagnetic centers of free radicals in biological materials. From received results registered EPR spectra it makes the interpretation of resonance curve on the basis of the intensity of signal proceed from the centers of paramagnetic chemical structures finding in samples. From the analysis of EPR spectra its gets information about number of spin of free radicals, the proprieties of structural and dynamic substances. The most often taken fragments from the patient organism it divides on several part and cools in liquid nitrogen. Sometimes is necessary the use in the investigations of the biological materials spin traps or spin tags. In conducted measurements it matches experimentally so low concentration of tags or traps, to after their connected with the sample's chemical structure did not come to the rise of the effect of extension of the EPR line in the result of the replaceable affect between their particles. We estimate if risen radicals in the result of the reaction in vivo, they have the influence on decay or modification: proteins, lipids, carbohydrates, nucleoids, etc (Nowak, Nedoszytko, 2005). Utilizing the EPR spectroscopy it moves the investigations of teeth and osseous tissues in the qualification of the crystal lattice structure changing in the result of biological processes connected with accomplished transplants, radiation diseases, the treatment of tumours, on the content of minerals (Jackson *et al.*, 2001; Zasacka *et al.*, 2008; Guzik *et al.*, 2002). The analysis of EPR spectrum of radicals being in the walls of arteries it moves under the influence of arteriosclerosis, the disorders of blood supply, the disease of the circulation, the tumours (Guzik *et al.*, 2002). However the lens of eye it studies under in relation to the cataract and concurrent her different diseases which have the influence on the registered amount spins e.g.: of diabetes (Gosławski *et al.*, 2008). Except diseases, the investigations conducts in the aim of the qualification of products consumed which influence profitably and unfavourably on the reactions of radicals in the metabolism of the organism (Rokyta *et al.,* 2004).

On the basis of spectral lines curve absorptive it establishes the dosimetry of medicines and antioxidants in the therapy of fighting morbid radicals. The sequence of studies conducts in the aim of estimation, which chemical compounds in medicines have the unfavorable influence on the human metabolism and the absorbed through the organism dose of the radioactive bundle. Establish from worked out investigations, that Zeeman splitting factor g in values 1 - 6 the most often come from the ions of transitory metals: Fe, Cu, Cr, Ti, Mn, Ni, Ce, Co (Kamińska et al., 2011; Zawada, 2009).

In our experiment the investigation on tissue specimens from the vagina in patients with the disorders of the statistics of sexual organs was conducted. Our studies on the tissue of women is one of the first studies in this area. Moreover, the EPR method is used in monitoring of oxidative stress (Palmieri, Sblendorio, 2007). The aim was to the investigate of the practical role of Electron Paramagnetic Resonance in the identification of free radicals in human tissues. The following work is a part of doctoral thesis of T. Góra entitled "The analysis of vaginal tissue slices extracted from patients with pelvic organ prolapse by means of the paramagnetic resonance spectroscopy".

## MATERIAL AND METHOD

Before we will execute measurements in EPR apparatuses it should earlier prepare the materials of the substance studied.

Pelvic organ prolapse occur through displacement of the uterus and the vagina. We differentiate two instances of this disorder: lowering (descensus) and prolapse (prolapsus). The disorders are a result of an imbalance between the forces holding the organs in their normal position and the forces affecting those organs (Skręt, Skręt-Magierło, 2007).

The condition afflicts around 15% of women to the extent of impairing the quality of living. Throughout their whole lives, as many as 50% of women experience problems connected with the genital statics disorders, according to scientific reports (Skręt, Skręt-Magierło, 2007; Richter, Varner, 2006).

The forces affecting the genitals, leading to their lowering or even prolapsing in extreme cases, are linked with an increased intraabdominal pressure (abdominal press) during physical activity, defecation or pushing during the second stage of labor. The forces holding genitals in their normal position are a result of the supporting and suspensory system's role. The system consists of connective tissue proper and muscular tissue: striated muscle and smooth muscle. The supporting system is composed of muscles forming the deep perineal pouch and the pelvic diaphragm, together with fascia and ligaments. The deep perineal pouch contains the deep transverse perineal muscle. The pelvic diaphragm is composed of the levatorani – its pubococcygeus and ischiococcygeus parts, and the coccygeus muscle. The suspensory system is formed by parametria and pelvic ligaments. The parametria are a scaffold of sorts for the genitalia. They are composed of connective tissue, bundles of which radiate from the cervix, becoming the matrix for the fibers of smooth muscles, arteries, venous plexuses and nerve fibers. We differentiate lateral, central and posterior parametria. The strongest part of the connective tissue system is located between the plates of the broad ligament of the uterus. It runs from the lateral edge of the cervix and is called the cardinal ligament (Mackenrodt's ligament) (Skręt, Skręt-Magierło, 2007; Richter, Varner, 2006; Youi, Authier, 2009; Cadenas, Davie, 2000).

A really crucial factor deciding about the correct tension of the muscular elements of the pelvic fundus is the innervation. It is derived from two sources: the pudendal nerve and the motor branches of the S3 and S4 spinal nerves. The pudendal nerve originates in the abdominal part of the sacral plexus and gives the inferior rectal nerve off to the sphincter ani muscle and then it divides into the perineal nerve and the dorsal nerve of the clitoris. The perineal nerve leads its fibers to the anterior part of the levator ani and the urethral sphincter muscle. The motor fibers of the S3 and S4 segments innervate the posterior part of the levator ani and the sphincter ani (Skręt, Skręt-Magierło, 2007; Richter, Varner, 2006).

The genital statics disorders are a result of a weakening within the supporting and suspensory system of the uterus or excessive strain coming from the forces affecting the genital organ. Comorbidity of both those causes is possible. Under normal circumstances, the uterus retains some motility, the cervix does not cross the interspinous line and the vagina and the rectum are aligned almost horizontally, with their lumina closed by the intraabdominal pressure (Skręt, Skręt-Magierło, 2007; Richter, Varner, 2006)

Diagnosis of the genital statics disorders is made on the basis of clinical trial and the results of additional tests. Patients with genital statics disorders complain of: pain in the hypogastric and sacral regions, protrusion of the vagina and/or the cervix from the pudendal cleft and urination and defecation disorders of various types (Skręt, Skręt-Magierło, 2007; Richter, Varner, 2006).

Among genital statics disorders risk factors there are: postmenopausal age, prior spontaneous deliveries, obesity, ailments accompanied by persistent cough, persistent constipations, physical overstrain, abnormal build of the bony pelvis, abnormal build of the spine, post-traumatic injury to the pudendal and pelvic nerves, neurogenic injury to the muscles of the pelvic fundus and iatrogenic causes (e.g. post-hysterectomy state). Qualitative changes of collagen (decrease in the amount of type I collagen, increase in the amount of type III collagen) together with the elastin fibers fragmentation is what underlies the genital statics disorders on the molecular level. Genetic factors are speculated to have an influence on the occurrence of the disorder (Non authors listed, 1996).

In order to grade the clinical severity of the disorder three descriptive systems are used: the Baden-Walker System (from 1972), the POP-Q System (from 1996) and the revised New York System (from 2000). Nowadays, it is the POP-Q System, suggested by Bump, that is commonly used. It allows for accurate and objective evaluation of the severity of the genital statics disorders basing only on the clinical trial. The trial requires a description of the maximal genital prolapse. The criteria of the maximal genital prolapse evaluation are fulfilled if the tension of the bulging vaginal wall does not increase during the push, pulling the prolapsing organ does not cause its farther prolapse, the patient confirms the stage of the prolapse as its highest and the examination in the standing position confirms the stage of the prolapse defined in another position. The level of the hymen serves as the point of reference. Against that level the position of six additional anatomical points (so called defined points) is defined. Grading of the static is further simplified by measuring: the urogenital hiatus (GH), the perineal body (PB), the total length of the vagina (TVL).

POP-Q System

Stage 0 - no signs of prolapse

Stage I – the most distal prolapse is more than 1 cm above the level of the hymen;

Stage II – the most distal prolapse is between 1 cm above or below the hymen;

Stage III – the most distal prolapse is more than 1 cm below the hymen but no further than (TVL-2) cm;

Stage IV – vaginal eversion or procidentia. The most distal prolapse protrudes to at least (TVL-2) cm against the level of the hymen (Non authors listed, 1996);

Among additional tests that help diagnose and define the causes of genital statics disorders there are: manometric techniques including intravaginal pressure evaluation, electrodiagnostic tests (electromyography and percutaneous stimulation of the perineal nerve) and imaging techniques (cystography, proctography, radiograph of the spine and the pelvis). The latest achievements in diagnostic techniques include the use of the paramagnetic resonance spectroscopy to measure the level of the oxygen free radicals. There are expectations for this technique to help understand the patomechanism of the disorder itself (Minini, Zanelli, 2008).

The oxygen free radicals, also called the reactive oxygen species (ROS), are oxygen molecules that carry a free unpaired electron. One of their main characteristics is their high chemical reactivity. Their biological activity involves muscle contractions, hormone secretion and the bacteriocidal effect of saliva, among others. The increase in the level of ROS in the human body comes as a result of a so called oxygen shock which occurs in the inflammation-ridden tissues. The excess of ROS has a toxic effect on the tissues (Youi, Authier, 2009; Cadenas, Davie, 2000). This effect has been confirmed primarily in the academic works on the circulatory system disorders and it involved destruction of the intercellular matrix structure, among others, including collagen and elastin (Zhao et al., 1987; Chen, Wen, 2005; Moalli et al., 2004; Moon et al., 2011; Drewes et al., 2007; Rhan et al., 2009). The processes of regeneration and destruction of elements of the intercellular matrix are closely connected to the activity of the matrix metalloproteinases 1-9 and their inhibitors (Vulic et al., 2011; Liang et al., 2010, Dyiri et al., 2011; Pierce et al., 2011; Chen et al., 2010; Wu, 2010).

Treatment of the genital statics disorders can be divided into conservative and operational. Conservative methods include pharmacotherapy (estrogen in balm or vaginal suppository) and physiotherapy (exercise for pelvic fundus muscles, electrostimulation) (Silva, Karram, 2006).

The operational treatment of the genital statics disorders includes correction of the suspensory system, entering through the vagina or the abdomen, or the correction of the supporting system by means of plastic surgery, using either the classic method or meshes (Skręt, Skręt-Magierło, 2007).

The samples were frozen immediately after surgery. They were stored for a short period of time. Preparation of the frozen samples was also performed in a short time. Weight for each sample was the same. Samples were placed in narrow test-glasses because of received mass 0.01 g. For the slowdown of free radical reactions all studied materials were frozen and was kept in the temperature 260 K. From organizational regards each sample received their own individual ticket. Samples were marked in the following way: with acids (bk), without acids (bbk), control with acids (bkk), control without acids (bkbk).

Analysing the spectrum we got the field of the surface in the twofold way: first through integration of the signal in OriginPro, second directly from the parameters of line ( intensity $2Y_{max}$ and the line width $\Delta Xpp$).

For the EPR measurements the standard X-band (~ 9 GHz) spectrometer, with digital registration of the spectra was used. The temperature measurements were done using the digital temperature control system (BRUKER ER 4131VT), measurement it was made in – 13°C.

RESULTS AND DISCUSSION

For each of the patients was performed from two to several runs. The selected spectra for each type of sample show Fig.1.

After registering all the lines have been smoothed out in OriginPro. Then, on the basis of the smoothed line, was read the following parameters: peak-to-peak line width $\Delta Xpp$, g-factor, the intensity of the line $2Y_{max}$ and the resonance field. The way in which these parameters have been determined is sketched in Fig. 2.

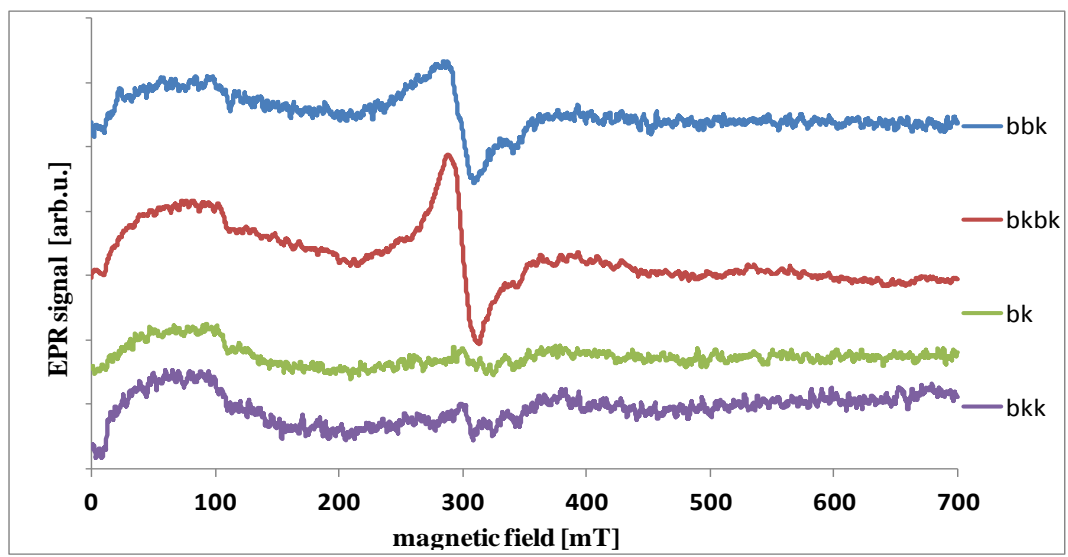

Figure 1. The EPR spectra of all kinds of samples.

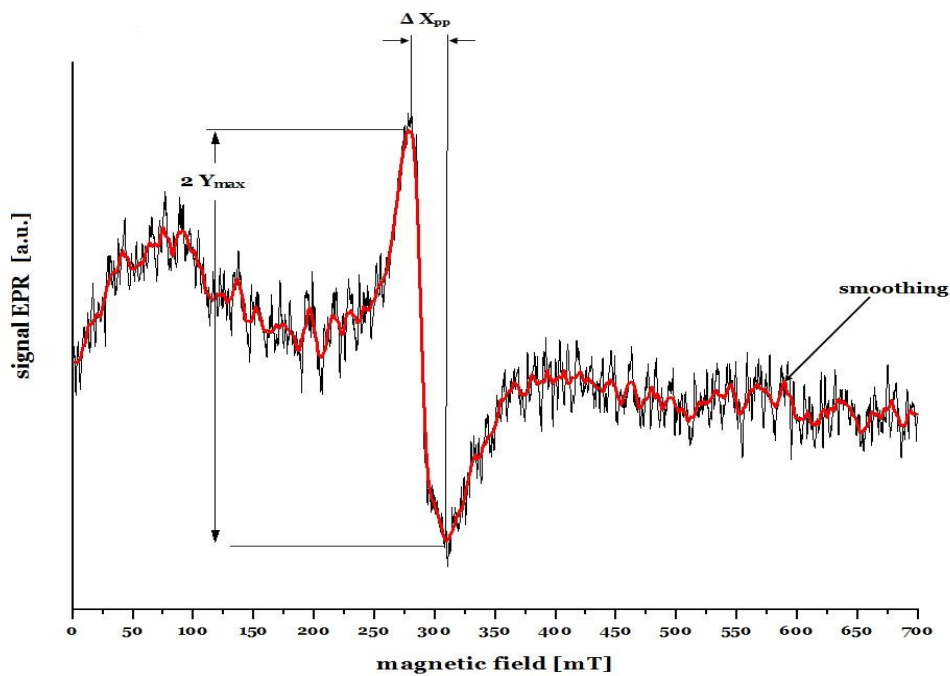

Figure 2. Examples of EPR spectra of the control sample (black line - the experimental spectrum, red line - smoothing out the program OriginPro.

These parameters were used to determine the amount of free radicals, based on the total intensity. It was determined in two ways:

a) directly from the spectrum - as it was showed on the Fig. 2 - from the peak to peak width and the intensity,

b) using the programme OriginPro-integrating smoothing out.

Taking into account the results from OriginPro we receive similar ones as in the direct method. Nevertheless, there does not change the relationship - peak without acids is stronger than this with acids. We receive results with certain difference in values; thi obtained directly from the spectrum (a) is 353078, however that received by use of the programme OriginPro-integrating (b) is 1326559.

Results were correlated between first and second method. From the example we chose the method of calculations directly for the spectrum. The signal after smoothing out was considerably better to the read of parameters, especially for the direct method.

The obtained values of total intensity for each series of measurements shows the Fig.3. While the averaged values of the two series of measurements shown in Fig. 4.

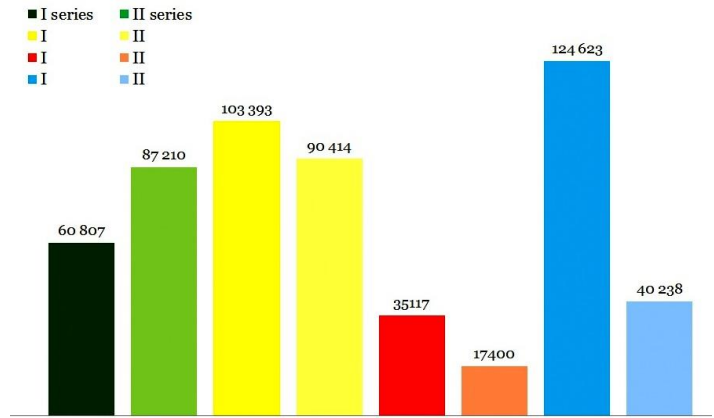

Figure 3. The mean value of integral intensity of EPR spectra for two measurement series.

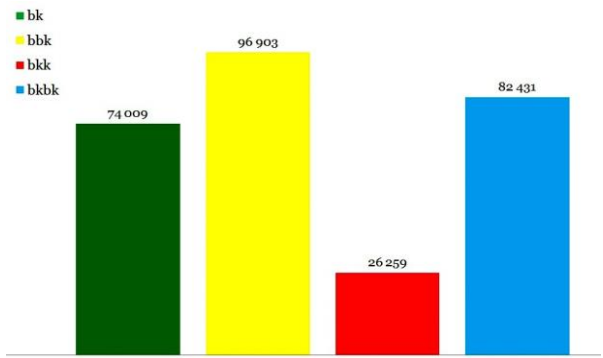

Figure 4. The mean value of integral intensity of EPR spectra for all kind of samples.

Based on the above graphs we observe a clear effect the consumption acid on female body.

We have determined from the EPR lines the for parameters: the peak – to – peak line width, the intensity as well as the resonance field. The source of the observed EPR signal could be the following: (1) centres with signals typical for protein peroxides, (2) centres with signals corresponding to ion free radicals associated with sulphur amino acids, (3) centers linked to organic complexes containing $Cu^{2+}$ (superoxide dismutase, meta proteinases) (4) centers associated with $Fe^{2+}$ and $Fe^{3+}$ ions present in hemoglobin, cytochromes and catalase (Guzik *et al.*, 2002).

The experimental results clearly show that the iron ions occur in the human tissue in two main forms, which give to different magnetic contributions well separated from each other: $Fe^{2+}$ ions of hemoglobin, which show the sharp EPR signal at g ~5.8 and give the paramagnetic contribution to the magnetization, and $Fe^{3+}$ ions of ferritin grains showing the broad EPR signal at g ~2 (Guzik *et al.*, 2002).

## CONCLUSION

Values appointed in the programme Origin have the ruthless value larger than from the direct methods. however, reflect exactly values got from the direct measurement.

As a result of the measurements observed a strong resemblance to studies (Guzik *et al.*, 2002). As in the previous work, we identified the ions $Fe^{2+}$ and $Fe^{3+}$.

In this work we observed a close relationship between the observed EPR signal and the consumption of omega 3 acids. The EPR seems to be a useful and valuable method for measuring presence of the free radicals in human tissue.